\documentclass[12pt]{iopart}

\newcommand{\nI} {{\mbox{Ni${}^I$}}}
\newcommand{\nII} {{\mbox{Ni$^{II}$}}}
\newcommand{\nIII} {{\mbox{Ni$^{III}$}}}
\newcommand{\nto} {{\mbox{Ni${}_3$TeO${}_6$}}}

\usepackage{graphicx}

\begin{document}

\title{$\nto$ -  a collinear antiferromagnet with ferromagnetic honeycomb planes}

\author{I \v Zivkovi\'c$^1$ $^2$, K Pr\v{s}a$^3$, O Zaharko$^3$ and H Berger$^4$}

\address{$^1$ Institute of Physics, Bijeni\v{c}ka c.~46, HR-10000, Zagreb, Croatia}
\address{$^2$ Laboratory for Quantum Magnetism, EPFL, CH-1015 Lausanne, Switzerland}
\address{$^3$ Lab. for Neutron Scattering, ETH Z\"urich and PSI, CH-5232 Villigen-PSI, Switzerland}
\address{$^4$ Institut de Physique de la Mati\`ere Complexe, EPFL, CH-1015 Lausanne, Switzerland}

\ead{zivkovic@ifs.hr}

\begin{abstract}
We report a comprehensive study of magnetic properties of $\nto$. The system crystallizes in a noncentrosymmetric rhombohedral lattice, space group R3. There are three differently coordinated Ni atoms in the unit cell. Two of them form an almost planar honeycomb lattice, while the third one is placed between the layers. Magnetization and specific heat measurements revealed a single magnetic ordering at $T_N = 52$ K. Below $T_N$ the susceptibility with the magnetic field parallel to the $c$ axis drops towards zero while the perpendicular susceptibility remains constant, a characteristic of antiferromagnetic materials. Neutron diffraction confirmed that the system is antiferromagnet below $T_N$ with ferromagnetic $ab$ planes stacked antiferromagnetically along the $c$ axis. All Ni moments are in the $S = 1$ spin state and point along the $c$ axis.

\end{abstract}

\pacs{1.05.fm,65.40.Ba,75.30.Cr,75.50.Ee}



\section{Introduction}
\label{Intro}

Investigations of simple ferromagnets (FMs) and antiferromagnets (AFMs) flourished in 60s and 70s of the last century. With both theoretical and experimental advances a rather comprehensive knowledge has accumulated and is now a part of textbooks. It is often used in a description of more complicated systems in an effort to approach the complexities from the well-known ground.

The recent hot topic in magnetism are the spiral antiferromagnetic multiferroics~\cite{CheongKhomskii}. In the noncentrosymmetric crystal structures incommensurate order may be induced by symmetry allowed Dzyaloshinskii-Moriya interaction, and, provided enough coupling between the magnetic and lattice degrees of freedom, one may find multiferroicity. Therefore, inspection of the crystal lattice and possible interactions is an invaluable tool in search for new materials with interesting properties.

Recently, the reinvestigation of the crystal structure of trinickel-tellurium-hexaoxide Ni$_3$TeO$_6$ has been reported~\cite{Becker2006}. It improved the crystal lattice parameters established in the initial report by Newnham and Meagher~\cite{Newnham1967}. Little has been done since the original paper on the characterization of the magnetic properties of $\nto$ with the only result published by Zupan and colleagues~\cite{Zupan1971}. Using the ESR technique on the powdered sample they measured the associated $g$-factor to be 2.26, similar to other Ni$^{2+}$ compounds~\cite{Richards1965,Yamaguchi1999}. Their temperature interval was restricted to above 100 K where no sign of a magnetic ordering has been observed. From the Curie-Weiss (CW) behaviour they obtained the Curie temperature $\theta = - 34$ K, indicating that the system is AFM.

Given the fact that $\nto$ lacks the center of inversion, we thought that if the anisotropy of Ni ions is such that in the ordered state the moments are oriented parallel to the plane, it would produce chirality and possibly a (ferro)electric response. However, the measurements of the dielectric constant have not revealed any signature of the (ferro)electricity.

Here we present the detailed investigation of the basic magnetic properties of $\nto$ using the neutron diffraction, magnetization and specific heat measurements on powdered and single crystal samples. We have determined that the system enters the ordered AFM state below $T_N = 52$ K with magnetic moments pointing along the $c$ axis. The magnetic sublattice consists of ferromagnetic honeycomb planes with alternating spin direction along the $c$ axis. The magnetization and specific heat measurements in dc magnetic fields parallel and perpendicular to the easy axis revealed that $\nto$ behaves similar to other canonical AFM compounds.

%
%
%
%

\section{Experimental details}
\label{Details}

Single crystals of the compound $\nto$ were synthesized via chemical vapour transport reactions. The starting materials were NiO (Alfa Aesar 99\% ), CuO (Alfa Aesar 99.99\% ), TeO$_2$ (Acros 99\% ), and NiCl$_2$ (Alfa Aesar 99.9\% ) and the crystals were grown from the non-stoichiometric molar ratio NiO : CuO : TeO$_2$ : NiCl$_2$ = 4 : 1 : 3 : 1. The starting powder was mixed in an agate mortar and placed in a quartz ampoule, which was evacuated to 10$^{-5}$ Torr and sealed. The ampoule was heated slowly to 700 $^0$C in a tube furnace and held there for four days followed by slow cooling (50 $^0$C/h) to room temperature. The sintered powder was dark green and polyphasic and its phase composition was not analyzed. About 20 g of this polyphasic powder mixture was placed in a silica tube, which subsequently was evacuated (10$^{-5}$ Torr), and electronic grade HCl was added in sufficient quantity to be used as transport agent. The ampoule was placed in a two zone gradient furnace between 750 $^0$C - 600 $^0$C and after ten weeks two different compounds were observed as single crystals:
\begin{enumerate}
\item {In the center of the ampoule cubic crystals with a maximum size of 5x5x5 mm$^{-3}$ of dark green Cu doped [Ni30Te32O90Cl2.67][Ni4.48Cl15.78]~\cite{Johnsson2004}}
\item {A number of triclinic plates with a maximum size of 6x6x1 mm$^{-3}$ of dark green $\nto$ formed at the cold end.}
\end{enumerate}
The obtained samples of $\nto$ were checked to ensure that they have no appreciable amount of impurity phases by powder X-ray and neutron measurements (below the detection limit). Also, no paramagnetic Curie-like contributions in DC magnetization at low temperature have been observed, even at higher fields, indicating the purity of the crystal.

Magnetization measurements were performed using a Quantum Design superconducting quantum interference device (SQUID) magnetometer in the temperature range 2 -- 300 K and with fields up to 5 T. The search for the spin-flop transition was performed on a Quantum Design PPMS system up to 9 T. Specific heat was measured using a Quantum Design PPMS system with a relaxation technique in fields up to 9 T. Neutron powder diffraction data have been collected from a 5 g polycrystalline sample loaded in a vanadium can (diameter 8 mm) with neutron wavelength of 2.566 \AA\ in the temperature range 3.35 K--60 K on the DMC diffractometer at SINQ, Paul Scherrer Institute, Villigen, Switzerland.

A visualization software VESTA~\cite{Momma2008} has been used for displaying the crystal structure.

%
%
%
%

%
%
\begin{figure*}
\includegraphics[width=\textwidth]{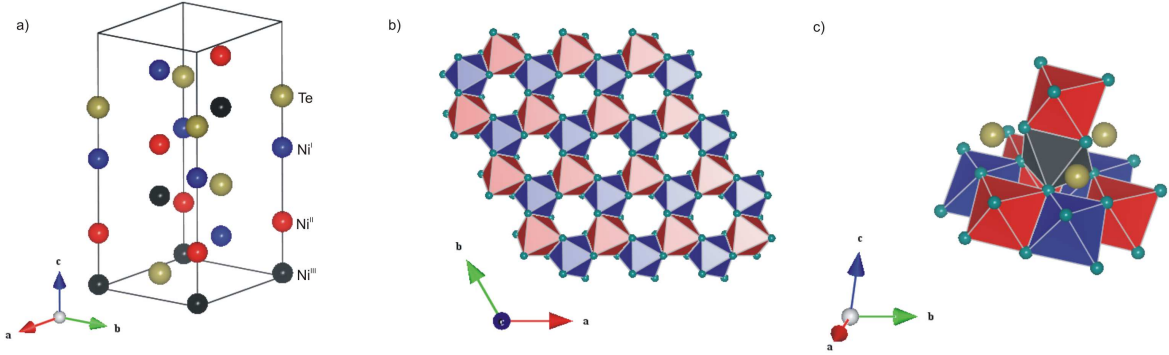}
\caption{(Colour online) Different views of the structure of $\nto$: a) the unit cell (oxygen ions are removed for clarity, $\nIII$ placed in the origin), b) view along the $c$ axis on the $ab$ plane with its hexagonal structure, c) the link between the $\nI - \nII$ hexagon (blue and red) and the $\nIII$ octahedron (black) along the $c$ axis. $\nII$ ion (red octahedron) on top of $\nIII$ belongs to the adjacent plane.}
\label{structure}
\end{figure*}

\begin{table}[b]
\caption{A summary of superexchange interactions in $\nto$. $d$ (A--B)\ is the distance in \AA\ between ions A and B. $N$\ is the number of superexchange connections between the pair of nickel ions.}
\begin{indented}
\item[]
\begin{tabular}{|c|c|c|c|c|c|}
\hline
Interaction & Atoms &  $d$(Ni--Ni) & Angle ($^0$) & $N$ & $d$(Ni--O)\\
\hline
\hline
$J_1$ (FM)		&	$\nI$-$\nII$			&	2.99		&	94.1		&	   2		&		1.99, 2.1 \\
$J_2$ (FM)		&	$\nII$-$\nIII$			&	2.81		&	83.9		&	   3		&		2.1, 2.1	\\
$J_3$ (AFM)	&		$\nIII$-$\nII$			&	3.44		&	120.2		&		1		&		1.99, 1.99 \\
$J_4$ (AFM)	&		$\nIII$-$\nI$			&	3.73		&	132.1		&		1		&		1.99, 2.1 \\
\hline
\end{tabular}
\end{indented}
\end{table}

\section{Results and discussion}
\label{Results}


A refinement of the crystal structure performed on the powder sample of $\nto$ confirmed the model suggested before~\cite{Newnham1967}. Here we recapitulate the main features which are important for the description of the magnetic behaviour of the compound. $\nto$ crystallizes in a rhombohedral lattice, space group R3 (no. 146). The unit cell parameters are $a = 5.103(2)$ \AA, $c = 13.755(10)$ \AA~ with Ni atoms occupying the $(0, 0, z)$ positions ($z_I = 0.352$, $z_{II} = 0.648$, $z_{III} = 0.852$).  Each Ni ion is surrounded with six oxygen ions that form a slightly distorted octahedron. The ligand environment is similar for all three Ni positions, although they differ in the coordination number.

As can be seen from the figure~\ref{structure}a, Te and Ni ions are stacked along the $c$ axis in a regular fashion forming the columns Te - $\nI$ - $\nII$ - $\nIII$. The nonmagnetic Te ion creates holes in the magnetic sublattice, thus preventing the direct magnetic exchange between $\nIII$ and $\nI$ moments.

Horizontally, $\nI$ and $\nII$ ions are connected through two oxygen ions and form an almost planar honeycomb lattice of edge-sharing octahedra (figure~\ref{structure}b). $\nI$ and $\nII$ ions are shifted slightly with respect to the plane in opposite directions. The $\nI - O - \nII$ angles are 94.1, suggesting the FM in-plane coupling ($J_1$). 

Each $\nII$ ion is directly linked with one $\nIII$ ion along the $c$ axis (figure~\ref{structure}c). There are three oxygens in between, all forming an angle $\nII - O - \nIII = 83.9^0$, again indicating the FM interaction ($J_2$). Two neighbouring planes are shifted relative to each other by $(\frac{a}{3},\frac{a}{3})$ so that each $\nIII$ ion is positioned above the center of the hexagon formed by $\nI$ and $\nII$ octahedra. There are two different angles between $\nIII$ and the members of the hexagon,  $\nIII - O - \nII = 120.2^0$ ($J_3$) and $\nIII - O - \nI = 132.1^0$ ($J_4$) which indicates that the inter-plane coupling should be AFM. However, the latter interaction also occurs between the $\nIII$ and $\nI$ of the same spin orientation and may lead to frustration (FM-FM-AFM triangles) in this material. We summarize all the anticipated interactions in table 1.

%
%
\begin{figure}[b]
\center
\includegraphics[width=0.6\textwidth]{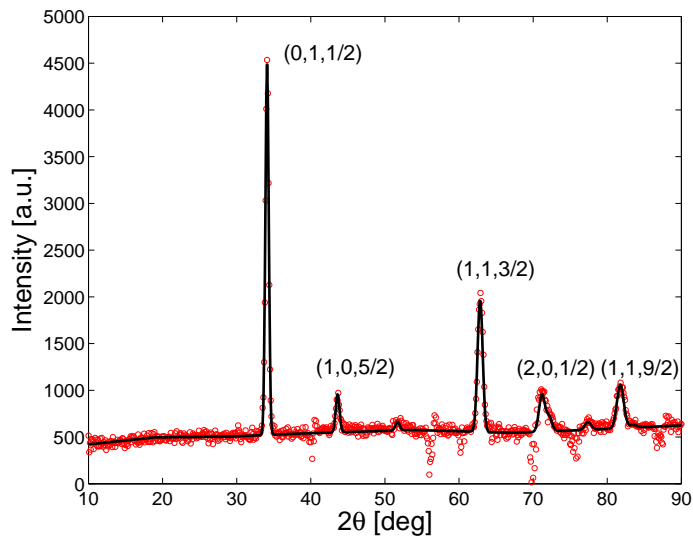}
\caption{(Color online) Observed (red dots), calculated (black line) and difference (blue line) neutron powder diffraction pattern of $\nto$ at 3.25 K - 60 K.}
\label{powder}
\end{figure}

All the observed peaks in the diffraction pattern could be indexed with the commensurate magnetic wave vector $\vec k=(0,0,1/2)$. The symmetry analysis for this wavevector reveals that the magnetic representation for all three Ni$^{2+}$\ sites can be decomposed into three irreducible onedimensional representations as $\Gamma=\Gamma_1\oplus \Gamma_2\oplus \Gamma_3$, summarized in table 2. Clearly, only the magnetic structures corresponding to irreducible representations $\Gamma_1$\ (giving all the moments along $c$) and $\Gamma_2\oplus \Gamma_3$\ (giving all the moments in the $ab$ plane) are possible. Only the symmetry-adapted mode belonging to $\Gamma_1$ fits to the experimental data.

\begin{table}
\caption{The Fourier coefficients of magnetic moments allowed according to the irreducible representations.}
\begin{indented}
\item[]
\begin{tabular}{|c|c|c|c|}
\hline
Site  &  $\Gamma_1$ & $\Gamma_2$ & $\Gamma_3$ \\
\hline
\hline
	$\nI$	& $(0,0,u)$		&	$(\frac{3-i\sqrt{3}}{2}u,-i\sqrt{3}u,0)$ &	$(\frac{3+i\sqrt{3}}{2}u,i\sqrt{3}u,0)$	 \\
	$\nII$	& $(0,0,v)$		&	$(\frac{3-i\sqrt{3}}{2}v,-i\sqrt{3}v,0)$ &	$(\frac{3+i\sqrt{3}}{2}v,i\sqrt{3}v,0)$	 \\
	$\nIII$	& $(0,0,w)$		&	$(\frac{3-i\sqrt{3}}{2}w,-i\sqrt{3}w,0)$ &	$(\frac{3+i\sqrt{3}}{2}w,i\sqrt{3}w,0)$	 \\
\hline
\end{tabular}
\end{indented}
\end{table}
 
%
%
\begin{figure}[b]
\center
\includegraphics[width=0.3\textwidth]{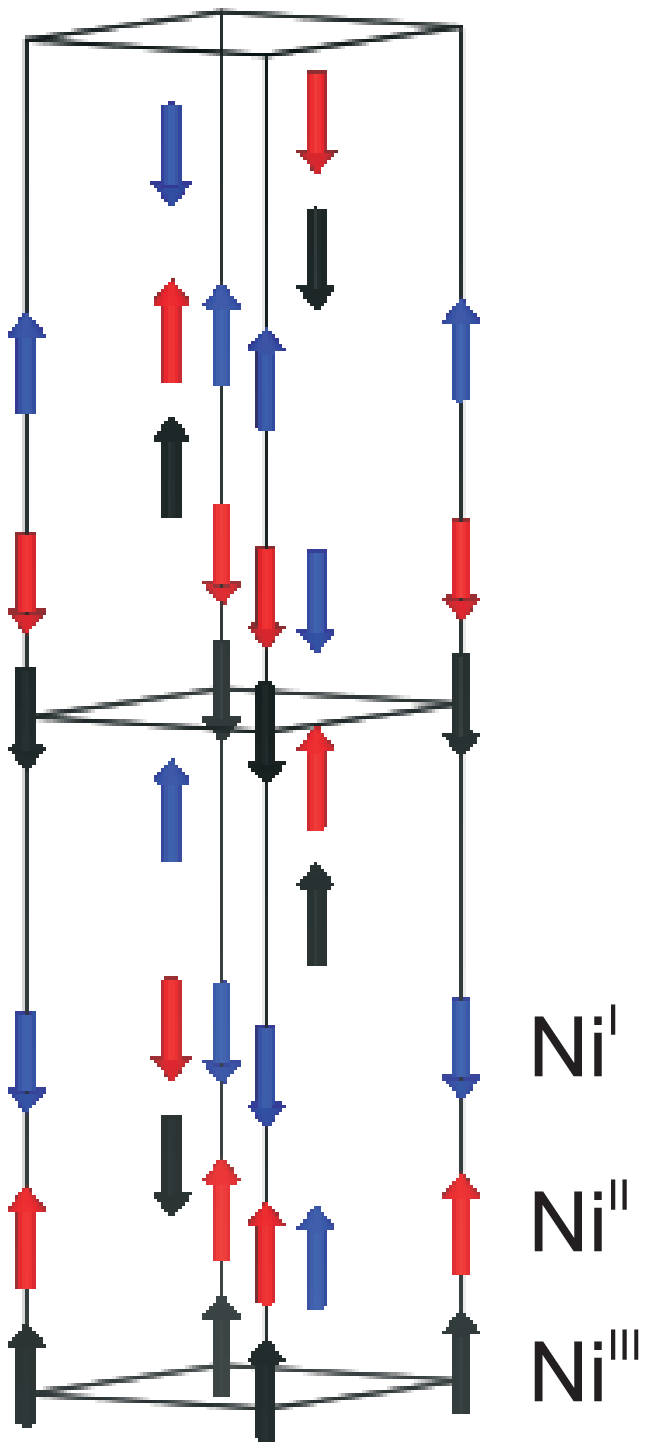}
\caption{(Colour online) The magnetic structure of $\nto$ deduced from the neutron powder diffraction. Arrows denote magnetic moments on particular nickel sites: \nI -- blue, \nII -- red and \nIII -- black.}
\label{magnetism}
\end{figure}

In figure~\ref{powder} we show the low temperature diffraction pattern, alongside the calculated profile for the magnetic structure presented in figure~\ref{magnetism}. A rather good match was obtained by assuming that the magnetic moments on all Ni$^{2+}$ ions are equal. This assumption is justified by the fact that although they don't have identical cation neighbours, their ligand environment (6 oxygen ions) is similar. The refinement (R$_F$=8.83\%) gave the value of $2.03(2) \mu_B/$\ ion, in accord with the spin value of $S = 1$ for Ni$^{2+}$\ ions.

The obtained magnetic structure is in excellent agreement with the conclusions drawn from the angles between the magnetic ions. $ab$ planes are ferromagnetic as well as the coupling between $\nII$ and $\nIII$ moments which sit on top of each other. The overall antiferromagnetic ground state is the result of the AFM interaction between $\nIII$ and the ferromagnetically coupled hexagon formed by three $\nI$ and three $\nII$ moments. We conclude that the antiferromagnetic $J_4$\ exchange interaction is not strong enough to cause the incommensurability of this magnetic structure.

\begin{figure}
\center
\includegraphics[width=0.6\textwidth]{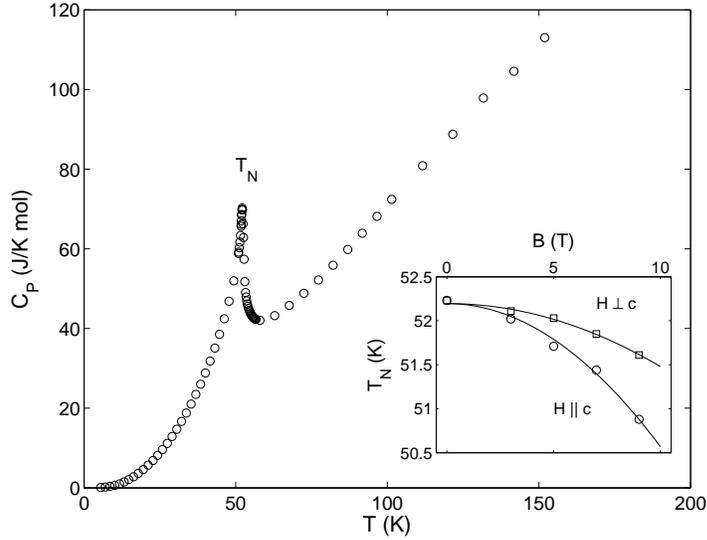}
\caption{The temperature dependence of the specific heat of $\nto$. Inset shows the field dependence of the transition temperature $T_N$ for the parallel and the perpendicular configuration. Solid lines represent fits to the quadratic dependence (see the text).}
\label{Cp}
\end{figure}

Specific heat has been measured with a dc magnetic field parallel and perpendicular to the $c$ axis. In figure~\ref{Cp} we show the temperature dependence of the specific heat in $H = 0$ T. Around 52 K there is a $\lambda$-like feature which marks the transition into a magnetically ordered state. When measured in the applied magnetic field, the transition shifts towards lower temperatures, as indicated in the inset of figure~\ref{Cp}. The shift is well reproduced by the quadratic dependence on the field
\begin{equation}
\nonumber
\delta T = T_N(H) - T_N (H=0) = - \alpha H^2
\end{equation}
where $T_N(H=0) = 52.20 \pm 0.01$ K is the temperature where $C_P$ has the maximum for zero field. In a simple molecular field approximation (MFA), the quadratic dependence was found for an uniaxial antiferromagnets~\cite{Heller1966} with a ratio $\alpha_{\parallel} / \alpha_{\perp} \approx 3$. From our measurements we find $\alpha_{\parallel} = 0.0163$ K/T and $\alpha_{\perp} = 0.0071$ K/T, giving a ratio $\alpha_{\parallel} / \alpha_{\perp} = 2.3$, fairly close to the MFA prediction.
%
%

\begin{figure}
\center
\includegraphics[width=0.6\textwidth]{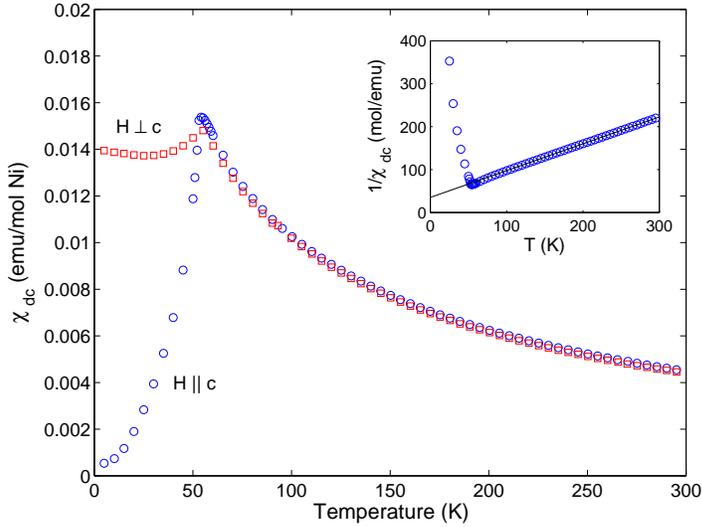}
\caption{(Colour online) The temperature dependence of the dc susceptibility of $\nto$ for the parallel and the perpendicular configuration with $H = 1000$ Oe. The inverse susceptibility for $H \parallel c$ with the Curie-Weiss fit is shown in the inset.}
\label{M-T}
\end{figure}

The temperature dependence of the dc susceptibility $\chi_{DC} = M/H$ for two orthogonal field directions is shown in figure~\ref{M-T}. At high temperatures both curves nicely follow the CW behaviour $\chi = C/(T + \theta)$ where $C$ is the Curie constant and $\theta$ is the Weiss temperature. Around 52 K the system orders and two curves show substantially different temperature dependence. For $H \parallel c$ the susceptibility drops quickly towards zero as the temperature is decreased. On the other hand, $H \perp c$ curve initially drops down but it levels off at low temperature, with a small minimum around 25 K. All these features are well-known characteristics of antiferromagnetic materials with moments pointing along the $c$ axis, in agreement with the magnetic structure deduced from the neutron diffraction.

Fitting the measured susceptibility above 150 K to the CW law, we obtain slightly different values for two field orientations: $C_{\parallel} = 1.599$ emu K/mol, $\theta_{\parallel} = 56.1$ K and $C_{\perp} = 1.545$ emu K/mol, $\theta_{\perp} = 51.6$ K. This is not unusual for antiferromagnetic compounds and has been explained in the case of MnF$_2$ to be due to the long-range dipole-dipole interaction~\cite{Okazaki1964,Nordblad1979}.

In this fit we have disregarded the temperature-independent contributions, namely the positive Van Vleck ($\chi_{vv}$) and the negative diamagnetic susceptibility ($\chi_d$). $\chi_{vv}$ can be calculated from
\begin{equation}
\nonumber
\chi_{vv} = \frac{8 \mu_B N_A}{\Delta} = 5.2 \cdot 10^{-7} \, \textrm{emu/mol}
\end{equation}
where $\mu_B$ is the Bohr magneton, $N_A$ is the Avogadro number and $\Delta = 10000$ cm$^{-1}$ is the energy gap between the octahedrally split $t_{2g}$ and $e_g$ levels measured by ESR~\cite{Zupan1971}. The diamagnetic susceptibility is~\cite{Selwood1956} $\chi_d = -1.2 \cdot 10^{-4}$ emu/mol, more than two orders of magnitude smaller than the measured value at 300 K. Both contributions can be neglected in the first approximation.

%
%

The values for the Weiss constants are in accordance with the observed transition around 52 K, indicating no frustration present in this material and, therefore, $J_4\ll J_3$. From the Curie constant we can calculate the effective magnetic moment $\mu_{eff} = 3.58 \mu_B$ in the case of the field parallel to the $c$ axis. This is somewhat larger than the theoretically predicted value for the $S=1$ system $\mu_{eff}^{calc}=3.2 \mu_B$ calculated using the measured value~\cite{Zupan1971} $g=2.26$.

Knowing the exact temperature dependence of the parallel and the perpendicular susceptibilities below $T_N$ one can in principle calculate the values of the exchange constant(s) using the Kubo's spin wave theory~\cite{Kubo1952}. However, it has been shown~\cite{Trapp1963} that there are large discrepancies between the experimental values and the theoretical predictions (up to 20\%), even for a simple system as MnF$_2$.

The modeling of $\nto$ is even more complicated by the fact that there are at least three different coupling constants to be considered:
\begin{itemize}
\item{in-plane $J_1$ between $\nI$ and $\nII$ (FM)}
\item{out-of-plane $J_2$ between $\nII$ and $\nIII$ which sit on top of each other (FM)}
\item{out-of-plane $J_3$\ between $\nIII$ and the in-plane hexagon.}
\end{itemize}
In addition to that, the anisotropy constant $K$ is unknown at the present, although it should be equal for all the moments (at least in the first approximation). Finally, each nickel ion has a different coordination number $z$. 

Given the fact that our measurements were done with the error in the alignment of the crystal axes with respect to the direction of the magnetic field not better than 5$^0$, we find that the modeling would not be reliable and leave the determination of the exchange constants for future inelastic neutron scattering experiments.

Measuring the magnetization up to 9 T with the magnetic field parallel to the $c$ axis we have not observed the transition to a spin-flop state. In simple AFMs, when the dc magnetic field is applied parallel to the easy axis, above the characteristic field $H_{SF}$ moments are perpendicular to the easy axis but still retain the antiparallel configuration. The magnitude of $H_{SF}$ depends on the anisotropy energy and the exchange energy, $\sqrt{2H_A H_E}$. In $\nto$, given the fact that there are at least three different $J$s, there is a possibility that more than one characteristic field is present for $H \parallel c$. Studies in larger magnetic fields would be desirable to elucidate this issue.

%
%
%
%
%

\section{Conclusions}
\label{Conclusion}

We have presented the results of the investigation of the magnetic properties of $\nto$. This material shows a single magnetic transition at $T_N = 52$\ K with a well defined $\lambda$-anomaly in the specific heat. Although the crystal structure indicates some magnetic frustration, below $T_N$ the system is a collinear antiferromagnet with ferromagnetically ordered $ab$\ honeycomb planes. The parallel and perpendicular susceptibilities below $T_N$\ display a canonical AFM behaviour with $\chi_{\parallel}$ reducing to zero for $T \rightarrow 0$ K. The spin-flop transition has not been observed up to 9 T. The preliminary dielectric constant $\epsilon_r(T)$ measurements down to 10 K do not show any features indicative of a dielectric transition.

\ack

We greatfully acknowledge T. Ivek for the experimental input. This work has been supported by the Croatian Ministry of Science, Education and Sports, project number 0035-0352843-2845 and Swiss National Science Foundation, SCOPES Project No. 111105. The sample preparation was supported by the NCCR research pool MaNEP of the Swiss NSF. This work was partially performed at SINQ, Paul Scherrer Institute, Villigen, Switzerland.

\end{document}